\documentclass[manuscript]{aastex}

\usepackage{txfonts}

\usepackage{epstopdf}

\usepackage{longtable}

\usepackage{natbib}

\slugcomment{For submission to the Astrophysical Journal}

\shorttitle{coronal heating}
\shortauthors{Bradshaw and Klimchuk}

\begin{document}

\title{Diagnosing the time-dependence of active region core heating from the emission measure: I. Low-frequency nanoflares}

\author{S. J. Bradshaw}
\affil{Department of Physics and Astronomy, Rice University, Houston, TX 77005, USA}
\email{stephen.bradshaw@rice.edu}\and
\author{J. A. Klimchuk}
\affil{NASA Goddard Space Flight Center, Solar Physics Lab., Code 671, 8800 Greenbelt Road, Greenbelt, MD 20771, USA}
\email{james.a.klimchuk@nasa.gov}\and
\author{J. W. Reep}
\affil{Department of Physics and Astronomy, Rice University, Houston, TX 77005, USA}
\email{jeffrey.reep@rice.edu}

\begin{abstract}
Observational measurements of active region emission measures contain clues to the time-dependence of the underlying heating mechanism. A strongly non-linear scaling of the emission measure with temperature indicates a large amount of hot plasma relative to warm plasma. A weakly non-linear (or linear) scaling of the emission measure indicates a relatively large amount of warm plasma, suggesting that the hot active region plasma is allowed to cool and so the heating is impulsive with a long repeat time. This case is called {\it low-frequency} nanoflare heating and we investigate its feasibility as an active region heating scenario here. We explore a parameter space of heating and coronal loop properties with a hydrodynamic model. For each model run, we calculate the slope $\alpha$ of the emission measure distribution $EM(T) \propto T^\alpha$. Our conclusions are: (1) low-frequency nanoflare heating is consistent with about 36\% of observed active region cores when uncertainties in the atomic data are not accounted for; (2) proper consideration of uncertainties yields a range in which as many as 77\% of observed active regions are consistent with low-frequency nanoflare heating and as few as zero; (3) low-frequency nanoflare heating cannot explain observed slopes greater than 3; (4) the upper limit to the volumetric energy release is in the region of 50~erg~cm$^{-3}$ to avoid unphysical magnetic field strengths; (5) the heating timescale may be short for loops of total length less than 40~Mm to be consistent with the observed range of slopes; (6) predicted slopes are consistently steeper for longer loops.
\end{abstract}

\keywords{Sun:~corona}

\section{Introduction}
\label{introduction}

One of the most enduring problems in modern astrophysics is to explain how the million-degree solar corona is created and sustained. The corona is a highly non-uniform environment which reflects its magnetic structure and activity. For example, there are magnetically {\it open} and {\it closed} regions; in open regions the field lines extend great distances into the heliosphere where they may reconnect with planetary magnetic fields, and in closed regions magnetic loops, line-tied at the solar surface and illuminated by plasma at EUV and X-ray temperatures, are observed. Quiet Sun regions are associated with relatively weak magnetic fields, whereas active regions, which overlie the sunspots at the surface, have stronger magnetic fields and give rise to a great deal of dynamic and eruptive solar activity. The magnetic fields are strongest in active region cores, where temperatures can exceed 5~MK. Regions of weaker magnetic field are generally associated with lower temperatures and it therefore seems clear that the heating rate must be in some way related to the magnetic field. The most obvious connection is through heating the plasma via the release of magnetic energy. This might occur by gradually stressing the magnetic fields threading the corona, building up the energy stored in the field, until the energy is released by reconnection leading to direct heating, bulk motion, and particle acceleration, the latter two ultimately being thermalized via collisional processes \citep[e.g.][]{Parker_1988}, for example. Alternatively, Alfv\'{e}n waves propagating along the field lines may interact, leading to energy release and dissipation via resonant processes \citep[e.g.][]{McIntosh_2011}. \citet{Klimchuk_2006} provides a more detailed discussion of these and other possibilities.  Unfortunately, direct observational signatures of coronal heating are extremely difficult to detect \citep{Bradshaw_2006,Reale_2008} and so the actual mechanism which raises the coronal plasma to such tremendous temperatures has not yet been identified.

Nonetheless there are observational measurements that may provide information concerning the properties of the heating mechanism, such as the timescale over which the release and dissipation of energy must occur, which may in turn yield clues to the physical nature of the mechanism itself. One popular conception of coronal heating is to consider the manner in which a monolithic (as observed) coronal loop is heated; if one assumes that the loop is composed of many thermally isolated, sub-resolution filaments, then the question to be answered concerns how each individual filament is heated. A single filament is understood to be the thinnest magnetic flux tube with an isothermal cross-section. There are two broad possibilities for the time-dependence of the heating mechanism for each filament; either steady or impulsive. In the case of steady heating energy is released and dissipated at a more or less constant rate, and the mechanism operates for a period of time that is much longer than the cooling timescale. In the case of impulsive heating energy is released and dissipated on a timescale that is significantly shorter than the cooling timescale. In the former case the filament heats up and eventually reaches some new, hot, hydrostatic equilibrium. In the latter case the filament heats up and is then allowed to cool and drain. We will refer to impulsive heating events by the generic term {\it nanoflare}, which is understood to mean any mechanism that gives rise to the impulsive release and dissipation of energy.

An important parameter is the frequency with which nanoflares occur on the same filament. For example, if a filament is heated and then cools and drains, returning to its initial state before being re-heated, then we refer to this as {\it low-frequency} nanoflare heating. However, if a filament is heated and then re-heated just a short while into its cooling and draining phase, then we refer to this as {\it high-frequency} nanoflare heating. We note that steady heating is just the upper limit of high-frequency nanoflare heating, where the delay between successive nanoflares on a single filament is effectively zero.

There has been a great deal of recent interest in one potential diagnostic of the frequency of occurrence of heating events in active region cores: the gradient or slope ($\alpha$) of the emission measure (in $\log-\log$ space) between the temperature of peak emission measure and 1~MK \citep{Warren_2011,Winebarger_2011,Tripathi_2011,Warren_2012,Schmelz_2012}. Of the active region cores analysed to date the observationally measured slopes are in the range $1.70 \le \alpha \le 5.17$. Steeper slopes indicate a greater proportion of hot (e.g. $3-5$~MK) material relative to warm ($\approx 1$~MK) material and this is consistent with steady heating (truly steady or high-frequency nanoflare heating). In this scenario most of the sub-resolution filaments are maintained at high temperatures, with relatively few cooling and draining; consequently there is little warm material. Shallower slopes indicate commensurately more equal proportions of hot and warm material, consistent with low-frequency heating. In this scenario the filaments are allowed to cool, with the result that there is more warm material than in the steady heating limit.

Recent work has focused on determining whether low-frequency nanoflares are consistent with the observed emission measure slopes and what the properties of the individual nanoflares (e.g. volumetric heating rate, duration) heating each filament must be, as well as the properties of the loops themselves (e.g. initial conditions, length). \cite{Warren_2011} found that neither low-frequency nanoflares nor steady heating could explain their observed emission measure slopes ($\approx 3.26$), though later observations yielded significantly shallower slopes \citep[$2.05-2.70$:][]{Tripathi_2011} which were more consistent with low-frequency heating. Steady heating led to effectively isothermal emission measures which severely underestimated the amount of warm plasma present, especially in diffuse regions where no discrete loops are observed. \cite{Mulu-Moore_2011} carried out several numerical experiments to study the emission measure slope in the case of heating by low-frequency nanoflares, focusing in particular on the influence of the radiative losses by using different abundance sets in their models. They found slopes in the range $1.60 \le \alpha \le 2.00$ for photospheric abundances and $2.00 \le \alpha \le 2.30$ for a coronal abundance set with an enhanced population of low first
ionization potential (FIP) elements (e.g. iron). \cite{Reale_2012} have also discussed the influence of radiative losses on the emission measure slope; using the most recent atomic data they found steeper emission measure slopes below 2~MK, which they associated with enhanced cooling in the $1-2$~MK range.

In the present work we focus on the heating of active region cores by low-frequency nanoflares and carry out an extensive survey of the parameter space of possible heating events and loop properties. In particular, we consider the importance to the emission measure slope of the volumetric heating rate, the duration of the heating $\tau_H$, the temporal envelope of the heating profile, the initial conditions in the loop ($n$, $T$), and the loop length. We carry out extremely detailed forward modeling to produce synthetic spectra and use this to calculate the emission measure for each of our numerical experiments. In Section~\ref{numerical} we describe our modeling approach and the experiments there were carried out. We discuss our
results in Section~\ref{results} and, finally, present a summary of our key results and a number of conclusions concerning heating by low-frequency nanoflares in Section~\ref{summary}.

\section{Numerical model and experiments}
\label{numerical}

Our approach to carrying out the numerical experiments and our forward modeling procedure are documented in detail in \cite{Bradshaw_2011}. However, we will provide a brief summary and
then focus on what is new here. We have explored an extensive parameter space of coronal loop and low-frequency nanoflare properties that we believe may be representative of active region core heating, and we present the results from 45 model runs. A low-frequency nanoflare is defined such that: (a) the heating timescale is significantly shorter than the timescale required for the strand to reach a new hydrostatic equilibrium if the increased heat input had remained steady; and (b) the time between each individual heating event on a single strand must be significantly longer than the cooling timescale following heating.

The first five columns of Table~\ref{table1} summarize the key loop properties and heating parameter values for each run. The first column indicates the number assigned to each Run, which will be used to reference them as they are discussed in turn. The second column gives the loop length $2L$ defined as the distance between the transition region foot-points along the loop. The computational domain also includes a deep (many scale heights) chromosphere at $2\times10^4$ K that is attached to each transition region foot-point in order to provide a source of material to ablate into the corona upon heating and to maintain a stable atmosphere \citep{Peres_1982}. The third and fourth columns describe the properties of the impulsive heating event. $E_{H0}$ and $\tau_H$ are the peak volumetric heating rate and the duration of heating. The fifth column gives the total volumetric heating input $E_H$. Note that the temporal envelope of the heating profile can be square ($E_H=E_{H0} \times \tau_H$) or triangular ($E_H=0.5 \times E_{H0} \times \tau_H$). The first 14 entries in Table~\ref{table1} correspond to the 14 numerical experiments that form the basis of \cite{Bradshaw_2011} and the remainder were conducted to substantially broaden the parameter space.

\begin{longtable}{c c c c c c c c}
\caption{A Summary of the Numerical Experiments Relating to the Loop and Impulsive Heating Event Properties.}\\
\hline
Run~\# & $2L$ & $E_{H0}$ & $\tau_H$ & $E_H$ & $\log_{10}~T_{\mbox{peak}}$ & $\alpha_{\mbox{model}}$ & $\alpha_{\mbox{observed}}$ \\
 & [Mm] & [erg~cm$^{-3}$~s$^{-1}$] & [s] & [erg~cm$^{-3}$] & [K] & & \\
\hline
\endhead
 1 & 20 & 0.05 & 10 & 0.5 & 5.85 & - & - \\
 2 & 20 & 0.10 & 10 & 1.0 & 5.95 & - & - \\
 3 & 20 & 0.10 & 30 & 3.0 & 6.15 & 0.58 & 1.17 \\
 4 & 20 & 0.10 & 100 & 10 & 6.45 & 0.76 & 0.81 \\
 5 & 20 & 0.10 & 300 & 30 & 6.35 & 0.98 & 0.83 \\
 6 & 20 & 1.00 & 10 & 10 & 6.35 & 1.38 & 0.89 \\
 7 & 20 & 1.00 & 30 & 30 & 6.45 & 1.95 & 1.65 \\
 8 & 20 & 5.00 & 10 & 50 & 6.55 & 1.79 & 1.73 \\
 9 & 80 & 0.10 & 10 & 1.0 & 6.25 & 1.05 & 1.69 \\
10 & 80 & 0.10 & 30 & 3.0 & 6.45 & 1.21 & 1.29 \\
11 & 80 & 0.10 & 100 & 10 & 6.45 & 2.24 & 2.22 \\
12 & 80 & 0.10 & 300 & 30 & 6.65 & 1.90 & 1.85 \\
13 & 80 & 1.00 & 10 & 10 & 6.55 & 1.89 & 2.12 \\
14 & 80 & 1.00 & 30 & 30 & 6.65 & 2.01 & 1.80 \\
\hline
15 & 40 & 0.03 & 500 & 7.5 & 6.45 & 1.14 & 1.27\\
16 & 40 & 0.06 & 500 & 15 & 6.55 & 1.32 & 1.68\\
17 & 40 & 0.10 & 500 & 25 & 6.55 & 1.58 & 1.15\\
18 & 40 & 0.50 & 500 & 125 & 6.75 & 1.95 & 1.93\\
19 & 80 & 0.03 & 500 & 7.5 & 6.55 & 1.40 & 1.85\\
20 & 80 & 0.06 & 500 & 15 & 6.65 & 1.53 & 2.16\\
21 & 80 & 0.10 & 500 & 25 & 6.65 & 1.73 & 1.54\\
22 & 80 & 0.50 & 500 & 125 & 6.95 & 1.72 & 2.09\\
23 & 160 & 0.03 & 500 & 7.5 & 6.65 & 1.68 & 2.47\\
24 & 160 & 0.06 & 500 & 15 & 6.65 & 2.01 & 1.77\\
25 & 160 & 0.10 & 500 & 25 & 6.75 & 1.90 & 1.95\\
26 & 160 & 0.50 & 500 & 125 & 7.05 & 1.74 & 2.20\\
27 & 160 & 0.03 & 2000 & 30 & 6.75 & 1.97 & 2.00\\
28 & 160 & 0.06 & 2000 & 60 & 6.85 & 1.92 & 2.14\\
29 & 160 & 0.10 & 2000 & 100 & 7.15 & 1.62 & 2.23\\
30 & 160 & 0.50 & 2000 & 500 & 7.35 & 1.71 & 2.26\\
\hline
31 & 40 & 0.03 & 500 & 15 & 6.45 & 1.45 & 1.66\\
32 & 40 & 0.06 & 500 & 30 & 6.65 & 1.42 & 1.24\\
33 & 40 & 0.10 & 500 & 50 & 6.65 & 1.63 & 1.50\\
34 & 80 & 0.03 & 500 & 15 & 6.55 & 1.81 & 2.16\\
35 & 80 & 0.06 & 500 & 30 & 6.65 & 1.78 & 1.60\\
36 & 80 & 0.10 & 500 & 50 & 6.75 & 1.81 & 1.82\\
37 & 160 & 0.03 & 500 & 15 & 6.65 & 1.98 & 1.76\\
38 & 160 & 0.06 & 500 & 30 & 6.75 & 2.03 & 2.04\\
39 & 160 & 0.10 & 500 & 50 & 6.85 & 1.93 & 2.14\\
\hline
40 & 80 & 0.20 & 10 & 1.0 & 6.35 & 1.49 & 2.64\\
41 & 80 & 0.20 & 30 & 3.0 & 6.45 & 1.46 & 2.56\\
42 & 80 & 0.20 & 60 & 6.0 & 6.45 & 1.91 & 1.93\\
43 & 80 & 0.20 & 120 & 12 & 6.55 & 1.95 & 2.28\\
44 & 80 & 0.20 & 300 & 30 & 6.65 & 2.05 & 1.90\\
45 & 80 & 0.20 & 600 & 60 & 6.75 & 2.04 & 2.16\\
\hline
\label{table1}
\end{longtable}

We solve the one dimensional hydrodynamic equations appropriate for a single magnetic strand in the field-aligned direction. The diameter of the strand cross-section is assumed to be so small that many strands are visible along the line of sight for any observing instrument.  In the case of a discernable loop, the many strands must combine in the resolved volume to give rise to the observed loop structure. We adopt a multi-species approach by treating electrons and ions as separate fluids, and couple them via Coulomb collisions. We preferentially heat the electrons \citep[though see][for an alternative scenario]{Longcope_2010} and therefore expect that $T_e \neq T_i$ during the heating and conductive cooling phases, and that $T_e \approx T_i$ at the onset of radiative cooling. We assume quasi-neutrality ($n_e = n_i = n$) and current free ($v_e = v_i = v$) conditions.

The equations solved are formulated to describe the conservation of mass, momentum and energy. They include transport and compression, viscous stress, gravitational acceleration and potential energy, Coulomb collisions, thermal conduction, optically-thin radiation, and external energy input (heating). Collisions between like-species are frequent enough that the electron and ion equations of state are given by $p_e = k_B n T_e$ and $p_i = k_B n T_i$, and only the thermal component of the electron energy and the thermal plus kinetic components of the ion energy are considered. Viscous interactions are expected to become extremely important at the high temperatures reached by some of our experiments and so must be included in the strand physics \citep{Peres_1993}. The optically-thin radiation calculation accounts for the volumetric energy loss due both to line emission \citep[for coronal abundances:][]{Feldman_1992} and thermal bremsstrahlung \citep[Chianti v6:][]{Dere_1997,Dere_2009}. Runs~$1-14$ and $40-45$ account for the effect of non-equilibrium ionization of Ca and Fe on the radiative losses, but deviations from equilibrium are only significant at times when radiation plays a very minor role in the energy balance (e.g. during heating and conductive cooling) and so the effect on the plasma thermodynamics is small. Nonetheless, the effect on the emission spectrum can be significant at higher temperatures \citep[e.g. above the temperature of peak emission measure:][]{Bradshaw_2006,Reale_2008}. The spatial profile of the volumetric energy input (heating) is such that it is uniform along the magnetic field and we adopt one of two temporal envelopes: (a) constant heating for a period $\tau_H$ (Runs~$1-14$ and $31-39$); and (b) a triangular profile with a linear increase to the maximum heating rate at $\tau_H / 2$ and then a linear decrease to zero at $\tau_H$ (Runs~$15-30$ and $40-45$).

We consider only spatially uniform heating in our numerical experiments for two reasons: (a) we are wary of introducing an additional, unconstrained parameter into our study; and (b) we believe that the radiative and enthalpy-driven cooling and draining phase, which determines the emission measure, is relatively insensitive to the details of the spatial distribution of heat in the impulsive case. Our justification for (b) is based on the efficiency of thermal conduction at redistributing the energy released into the loop. In order to obtain emission measures that peak at temperatures ($T_{\mbox{peak}}$) that are consistent with the observed range, it is necessary to heat the strand to temperatures significantly greater than $T_{\mbox{peak}}$ (e.g. into the region of 10 MK). At these temperatures, and at the rarefied densities of the initial conditions, thermal conduction redistributes the energy throughout the loop in just a few seconds, which is effectively akin to uniform heating on the timescales of interest to us. By the time the strand enters the radiative cooling and draining phase then all `memory' of the initial spatial distribution of heat will have been lost \citep{Winebarger_2004,Patsourakos_2005}.

We use the numerical code HYDRAD \citep{Bradshaw_2003,Bradshaw_2011} to solve the two-fluid hydrodynamic equations under the circumstances described above. HYDRAD has several desirable features which make it ideal for application to the study of impulsive heating to extremely high temperatures. Written exclusively in C++, it is fast and robust, and models an entire loop strand (foot-point to foot-point for any geometry via an analytical equation, or look-up table, for gravity) with an adaptive grid for efficiently capturing small-scale properties of the solution. It is user-friendly and easily configurable via a Java-developed graphical user interface.

The forward modeling aspect of this work follows the procedure described by \cite{Bradshaw_2011}. Although the EUV emission from active region cores is generally diffuse \citep[only a fraction of the EUV emission is contained in observationally discernable loops:][]{Viall_2011}, we model the core as though it were a single multi-stranded loop. We constructed a snapshot of the loop from the data output by each experiment. The loop is comprised of sub-resolution magnetic strands, where each strand represents one stage (captured in 1 second intervals) of the heating and cooling cycle. Therefore, a loop with a heating and cooling cycle lasting 2000~s would be composed of 2000 individual strands \citep[see also][]{Guarrasi_2010}. We then calculated the emission measure in the region of the loop apex, since we are interested in active region core heating and wish to avoid foot-point~/~moss contamination, using two different methods. The first gives the 'true' or model emission measure, which is obtained directly from the numerical data ($n$ and $T$) stored in each grid cell. The second is the `observed' emission measure, forward modeled by applying the Pottasch method \citep{Pottasch_1963,Tripathi_2011} to a {\it Hinode}-EIS \citep{Culhane_2007} spectrum synthesized from the numerical data. The intensity of emission from each resolved volume element along the loop, recorded on the corresponding detector pixel, is calculated by summing the line-of-sight emission from each sub-resolution strand folded through the appropriate instrument response function \citep[e.g. Equations (8), (9) and Figure~1 of][]{Bradshaw_2011}. Note that the `observed' emission measure is idealized in that there are no errors due to photon statistics, photometric calibration of the instrument, or incorrect atomic physics (abundance, collision and excitation rates, etc.). The same atomic physics is used to generate the synthetic spectrum as is used to infer the emission measures from that spectrum, so any errors cancel out. \cite{Landi_2012} have investigated the effect of using inconsistent atomic physics in differential emission measure (DEM) diagnostics and found that it can lead to uncertainty in the peak DEM value. Uncertainties associated with real data are substantial and can have a significant impact on the emission measure slope, as we discuss later.

We performed this calculation for a number of individual spectral lines and applied the Pottasch method to each (as though they were observed lines) in order to derive an emission measure plot.  An example is shown in Figure~\ref{fig1} for Run 44. The plus signs show the ‘observed’ measures and the diamonds show the true values.  EM loci curves are also plotted for each line. These curves would intersect at a single temperature if the plasma were isothermal, and the degree of deviation from a single intersection point is an indication of the multi-thermality of the plasma \citep{Landi_2010}. We used a linear least-squares fit to calculate the slope of each emission measure distribution between the temperature of its peak and $10^6$~K. A steeper slope is consistent with a narrower, more isothermal, emission measure and a shallower slope with a broader, multi-thermal distribution of plasma. The spectral lines in the EIS wavelength channels used to construct the forward modeled emission measures used for our study are listed in Table~\ref{table2}.

\begin{longtable}{c c c}
\caption{A List of the Spectral Lines Used by the Forward Modeling Code to Generate the Emission Measure Loci Plots.}\\
\hline
Ion & Wavelength~(\AA) & $\log_{10}~T$~(K) \\
\hline
\endhead
 Mg~V & 276.579 & 5.45 \\
 Mg~VI & 268.991 & 5.65 \\
 Mg~VI & 270.391 & 5.65 \\
 Si~VII & 275.354 & 5.80 \\
 Mg~VII & 278.404 & 5.80 \\
 Mg~VII & 280.745 & 5.80 \\
 Fe~IX & 188.497 & 5.85 \\
 Fe~IX & 197.865 & 5.85 \\
 Si~IX & 258.082 & 6.05 \\
 Fe~X & 184.357 & 6.05 \\
 Fe~XI & 180.408 & 6.15 \\
 Fe~XI & 188.232 & 6.15 \\
 Si~X & 258.371 & 6.15 \\
 Si~X & 261.044 & 6.15 \\
 S~X & 264.231 & 6.15 \\
 Fe~XII & 192.394 & 6.20 \\
 Fe~XII & 195.119 & 6.20 \\
 Fe~XIII & 202.044 & 6.25 \\
 Fe~XIII & 203.828 & 6.25 \\
 Fe~XIV & 264.790 & 6.30 \\
 Fe~XIV & 270.522 & 6.30 \\
 Fe~XIV & 274.204 & 6.30 \\
 Fe~XV & 284.163 & 6.35 \\
 S~XIII & 256.685 & 6.40 \\
 Fe~XVI & 262.976 & 6.45 \\
 Ca~XIV & 193.866 & 6.55 \\
 Ca~XV & 200.972 & 6.65 \\
 Ca~XVI & 208.604 & 6.70 \\
 Ca~XVII & 192.853 & 6.75 \\
 Fe~XVII & 269.494 & 6.75 \\
\hline
\label{table2}
\end{longtable}

\section{Results}
\label{results}

The key results of our study are presented in the final three columns of Table~\ref{table1}. The sixth column of Table~\ref{table1} gives the temperature of the emission measure peak ($\log_{10}~T_{\mbox{peak}}$) and the final two columns give the slopes ($\alpha$) of the model and forward modeled emission measures for $6.0 \le \log_{10} T \le \log_{10} T_{\mbox{peak}}$, where $EM(T) \propto T^\alpha$. The first finding to which we draw attention are the differences between the slopes calculated for the model and the forward modeled `observed' emission measure. We find a general tendency for the slope in the forward modeled case to be steeper than the model slope as the loop length increases; it is true for 33\% of the 20~Mm loops, 43\% of the 40~Mm loops, 68\% of the 80~Mm loops, and 82\% of the 160~Mm loops in Table~\ref{table1}. This implies that the slope calculated from observationally derived
emission measures may not reflect the true slope, which would be the slope obtained if one were in possession of `perfect' data (e.g. in the model case this would be the actual grid cell densities and temperatures). Forward modeling then allows us to investigate how the biases introduced by commonly used observational tools and techniques may influence sets of results and the conclusions associated with them.

\begin{figure}
\includegraphics[width=0.8\textwidth]{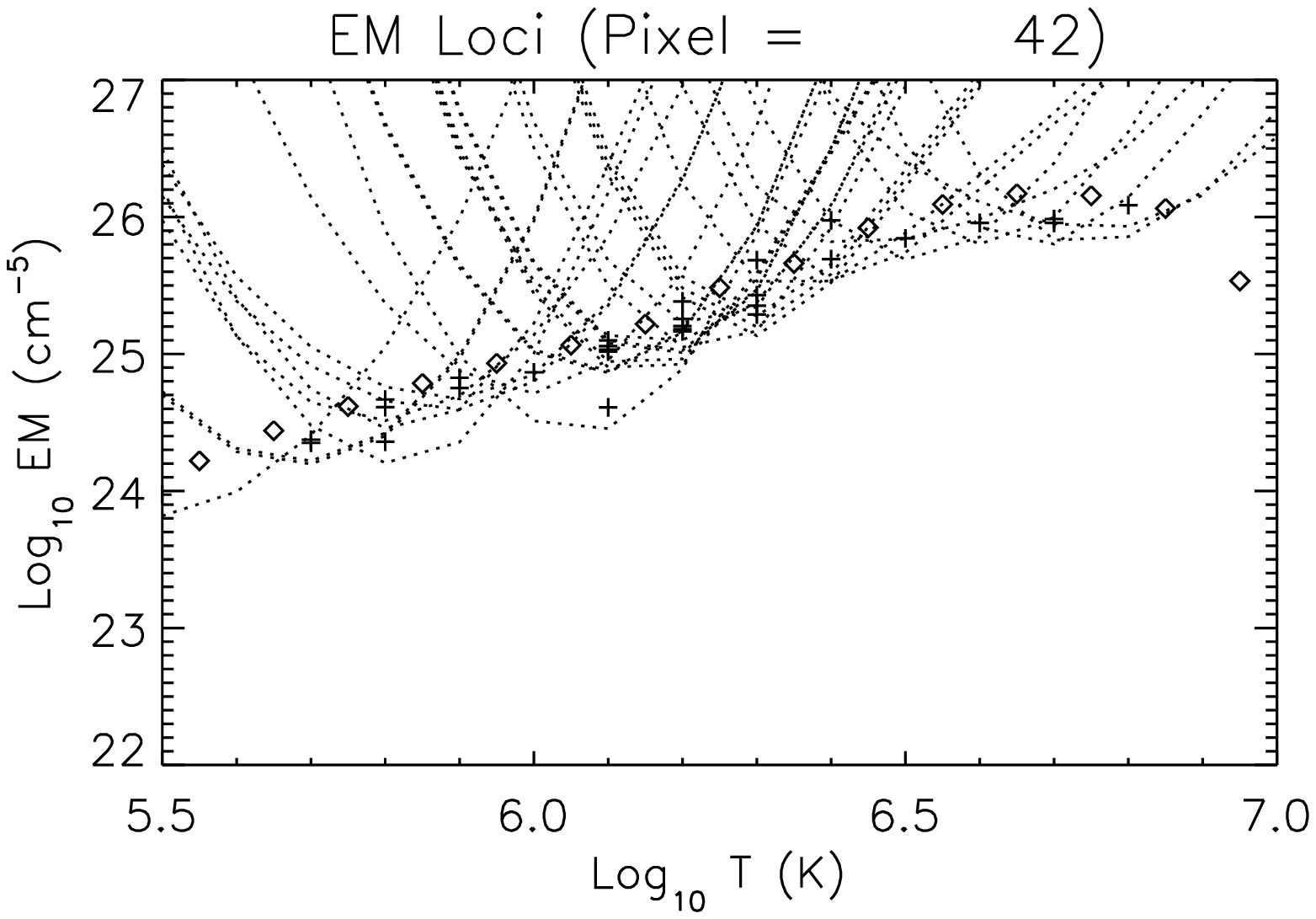}
\includegraphics[width=0.8\textwidth]{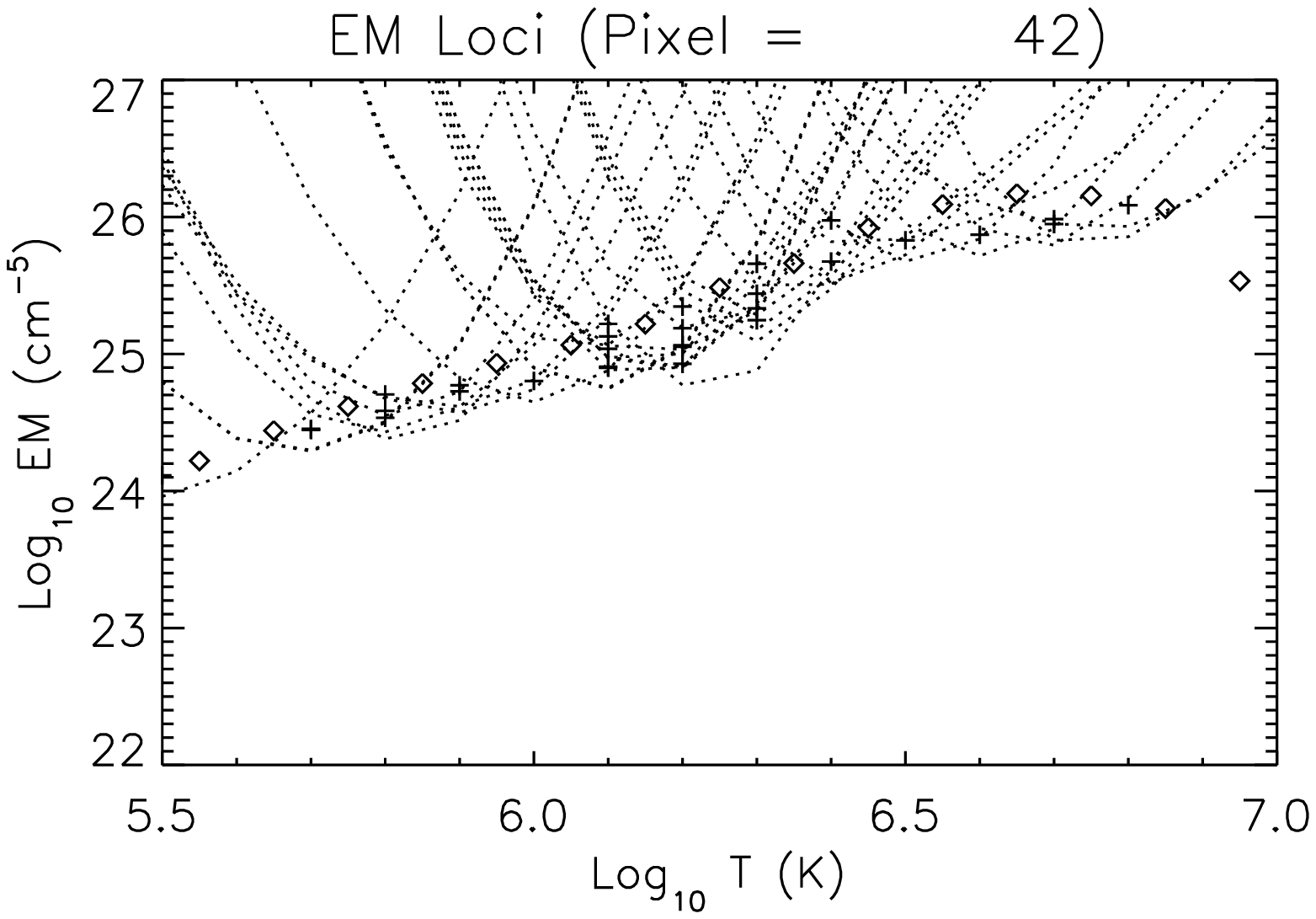}
\caption{Emission measure loci plots for Run 44. Diamonds are the `perfect' model emission measure. + signs are the emission measure values calculated by applying the Pottasch method to a forward modeled Hinode-EIS spectrum.  Top (bottom) plot assumes a density of $10^{10}$~cm$^{-3}$ ($10^{9}$~cm$^{-3}$) for the contribution functions. Pixel 42 is the loop apex pixel on the virtual detector.}
\label{fig1}
\end{figure}

By way of an example consider the Mg~VII 280.745~\AA~line, which \cite{Warren_2011} noted was not consistent with other lines formed at a similar temperature (e.g. Si~VII 275.354~\AA). They conjectured that since the Si~X lines in their study were consistent with the Fe lines then the issue must lie with the Mg abundance and it was multiplied by a factor of 1.7 to force agreement between the Mg~VII 280.745~\AA~and the Si~VII 275.354~\AA~lines. However the upper panel of Figure~\ref{fig1} shows an example emission measure loci plot, calculated using a spectrum forward modeled from numerical data, with two lines towards the low temperature range that are not consistent with their neighbours. The line formed at $10^{5.8}$~K is the Mg~VII line, but the line formed at $10^{6.05}$ is the Si~IX 258.082~\AA~line. Since the Si~IX line is not consistent with the neighbouring Si~X lines then the problem cannot be related to the element abundance. Instead, it is related to the density sensitivity of these emission lines. The emission measure loci plot in the upper panel of Figure~\ref{fig1} was calculated for a density of $10^{10}$~cm$^{-3}$, commonly assumed for an active region core \citep[e.g.][]{Tripathi_2011}, in the contribution function. The lower panel of Figure~\ref{fig1} shows the same emission measure loci plot calculated for a density of $10^9$~cm$^{-3}$ in the contribution function. The Mg and Si lines are brought into consistency with one another and with the lines that are not density sensitive (which do not change). In addition, there is less scatter between lines formed at the same temperature. The improved agreement
between the density sensitive and non-density sensitive lines is a strong indication that the lower density is a better estimate of the active region core density. The best value is probably somewhat greater, but closer to $10^9$~cm$^{-3}$.

In the current work we use a value of $10^{10}$~cm$^{-3}$ for consistency with \cite{Tripathi_2011} who adopt this value and also employ the Pottasch method for calculating their emission measures. Our aim is to be as consistent as possible with observational studies in our forward modeling. We note that using different densities in the contribution function will yield different emission measure slopes. The differences between the model and forward modeled slopes is of particular relevance to studies of emission measure slopes as diagnostics of the time-dependence of the underlying heating mechanism. Previous studies have found that emission measure slopes calculated from `perfect' model data are too shallow, in the case of low-frequency nanoflares, to explain observed slopes in excess of 3.

\begin{longtable}{c c c c}
\caption{A Summary of Recent Observational and Numerical Modeling Results Concerning the Slope of the Emission Measure Coolward of the Peak.}\\
\hline
 Slopes & Range of $\log_{10} T$ & Reference & Comments \\
\hline
\endhead
3.26 & $6.00-6.60$ & \cite{Warren_2011} & 1~AR; $10\arcsec \times 15\arcsec$~sub-region; \\
& & & MCMC method; with background \\
2.17, $\infty$ & & & Model slopes; \\
& & & low~\&~high~frequency nanoflares \\
\hline
3.20 & $6.00-6.50$ & \cite{Winebarger_2011} & 1~AR; $5\arcsec \times 25\arcsec$~sub-region; \\
& & & xrt\_dem\_interative2.pro; \\
& & & background subtract \\
\hline
$2.08-2.47$ & $5.50-6.55$ & \cite{Tripathi_2011} & with background \\
$2.05-2.70$ & & & background subtract \\
& & & 2~ARs; $5\arcsec \times 5\arcsec$~to~$10\arcsec \times 15\arcsec$~sub-regions \\
\hline
$1.60-2.00$ & $6.00-[6.60-6.80]$ & \cite{Mulu-Moore_2011} & Model slopes; photospheric abundances \\
$2.00-2.30$ & & & coronal abundances \\
\hline
$1.70-4.50$ & $6.00-6.60$ & \cite{Warren_2012} & $11 \ge 3.00$,~$5 \approx 2.00$; \\
& & & 12~ARs; 2~with~2~sub-regions; \\
& & & 1~with~3~sub-regions; multiple~pixels; \\
& & & MCMC method; with background; \\
& & & slope increases with unsigned flux \\
\hline
$1.91-5.17$ & $6.00-[6.30,6.80]$ & \cite{Schmelz_2012} & $4 < 2.60$,~$2 > 3.00$; \\
& & & 5~ARs; 2~with~2~sub-regions; \\
& & & 3~with~1~sub-region; multiple~pixels; \\
& & & xrt\_dem\_iterative2.pro, dem\_manual.pro; \\
& & & with background (subtract has small effect); \\
& & & slope increases with AR age \\
\hline
\label{table3a}
\end{longtable}

Table~\ref{table3a} provides a summary of observational measurements in active region cores and numerical calculations of the emission measure slope coolward of the peak. The column `Slopes' refers to the range of emission measure slopes coolward of the peak, `Range of $\log_{10} T$' gives the temperature range over which the slopes were calculated and `Reference' the paper in which the work is described. The `Comments' column discusses any particulars that should be considered when making comparisons between the different studies. The Markov-Chain Monte Carlo (MCMC) method is described by \cite{Kashyap_1998}, xrt\_dem\_iterative2.pro is a SolarSoft (SSW) routine used to generate emission measures constrained at high temperatures by
Hinode-XRT data, dem\_manual.pro is an alternative SSW routine for emission measure calculations, Pottasch refers to the method described by \cite{Pottasch_1963} and `Model slopes' indicates that the emission measure was calculated from `perfect' model data.

\begin{longtable}{c c c c c c}
\caption{Distribution of Observed Emission Measure Slopes Coolward of the Peak.}\\
\hline
 & $\alpha \le 2.00$ & $2.00 < \alpha \le 2.50$ & $2.50 < \alpha \le 3.00$ & $3.00 < \alpha \le 3.50$ & $\alpha > 3.50$ \\
\hline
\endhead
$\alpha$ & 3  & 5 & 3 & 6 & 5 \\
$\alpha - \Delta \alpha$ & 11 & 6 & 2 & 2 & 1 \\
$\alpha + \Delta \alpha$ &    &   & 3 & 5 & 14 \\
\hline 
\label{table3b}
\end{longtable}

Table~\ref{table3b} shows the distribution of the observed values of $\alpha$ in bins of width 0.5, where we have used the average value in cases where the same active region is observed on the same date. The first row assumes that there are no errors in the measurements. \cite{Guennou_2012} have recently performed a detailed analysis of the effects of realistic measurement errors due photon statistics and inaccurate atomic physics. Using a Monte Carlo approach combined with Bayesian statistics, they concluded that observationally derived emission measure slopes are uncertain by typically $\pm 1.0$.  Atomic physics uncertainties dominate. After consulting several spectroscopists, they assumed the following in their synthetic observations: 20\% random errors that are different for every spectral line (uncertain radiation and excitation rates and atomic structure calculations); 30\% random errors that are the same for every line from a given ion but different for different ions (uncertain ionization and recombination rates); 30\% random errors that are the same for every ion of a given element but different for different elements (uncertain abundances other than the first ionization potential (FIP) bias); and 30\% random errors that are the same for all low-FIP elements (the uncertain FIP bias). The second and third rows in Table~\ref{table3b} indicate the distribution of observed $\alpha$ under the assumption that the measurements are in error by $\Delta \alpha = +1$ and $-1$.

\cite{Mulu-Moore_2011} found maximum slopes of 2.3 from their models using coronal abundances and suggested that enhanced cooling due to an element population having super-coronal abundances in active region cores could give rise to yet steeper slopes. \cite{Reale_2012} demonstrated that new atomic calculations using more sophisticated models and including many more emission lines at lower coronal temperatures can yield new estimates of the radiative losses that are significantly enhanced, leading to correspondingly steeper emission measure slopes. Though the emission measure slopes calculated from forward modeled spectra can be greater than the model values in some cases, we note that for low-frequency nanoflares it is still difficult to obtain slopes that exceed 2.3. There are only three examples in Table~\ref{table1} (Runs~[23,40,41]), all for relatively low total volumetric heating ($<10$~erg~cm$^{-3}$). Nonetheless, forward modeling shows that the applicability of low-frequency nanoflares may extend somewhat further into the lower end of the range of observational results than may be expected from calculating the slope from `perfect' model data alone.

Our results encapsulate a parameter space within which we investigate variations in the loop length, the magnitude and timescale of energy release, and the temporal envelope of the energy release mechanism. We now discuss the influence of each parameter in turn, with reference to the results listed in Table~\ref{table1}.

We have chosen lengths such that $20 \le 2L \le 160$~Mm. Assuming approximately semi-circular loops then a total length of 160~Mm yields a foot-point separation of about 100~Mm or 140\arcsec, a reasonable upper limit for the diameter of an active region core. The general trend that we observe in our numerical results is for the model and forward modeled emission measure slopes to steepen with increasing $2L$. This becomes most apparent when comparing particular groups of Runs. For example, consider the pairs of Runs~[3,10], [4,11], and [5,12] for which only the length changes (we discard Run~2 and its corresponding Run~9 because the emission measure peaks below 1~MK in Run~2). The slope can be seen to steepen, but of these Runs only 11~\&~12 may be of interest because the slopes are consistent with the lower end of the observed range and $T_{\mbox{peak}}$ is consistent with the temperature range of peak active region core emission ($6.45 \le \log_{10} T_{\mbox{peak}} \le 6.75 $). Runs~$15-26$ and $31-39$ constitute a more detailed study of the relationship between the loop length and the slope. Again, considering equivalent triples of Runs where only the length changes (e.g. [15,19,23], [16,20,24], and [31,34,37], etc.) we see that the trend is largely maintained, though the slope of the forward modeled emission measure for Run~20 is somewhat steep in comparison with Runs~16~\&~24. Likewise Run~34. These anomalies
could likely be corrected by choosing a better estimate for the density in the contribution function, bringing the forward modeled slope into closer agreement with the model slope. Within these
groups of Runs a peak heating rate of 0.5~erg~cm$^{-3}$~s$^{-1}$ yields a $T_{\mbox{peak}}$ that is in general too high to be consistent with active region cores for the length range $L >40$~Mm. In most cases a peak heating rate of 0.1~erg~cm$^{-3}$~s$^{-1}$ yields $T_{\mbox{peak}}$ in the region of those observed \citep[$6.40 \le \log_{10} T \le 6.65$:][]{Warren_2012} and so the upper limit to the rate of energy release probably lies between 0.1 and 0.5~erg~cm$^{-3}$~s$^{-1}$ for timescales of O(100)~s (stronger heating may be permissible on shorter timescales, we consider this below). Our experiments indicate that, regardless of the loop length, low-frequency nanoflares can only account for the lower end of the range of observed slopes ($\le2.6$, assuming no errors) while remaining consistent with the peak emission temperature of active region cores.

We can understand the dependence of the emission measure slope on the loop length by considering the work of \cite{Bradshaw_2010} who showed that the value of the index in the scaling law $T \propto n^\delta$ depends on the loop length, with smaller values of $\delta$ for longer loops. $\delta$ itself depends on the ratio of the draining timescale to the radiative cooling timescale in the corona. Longer loops exhibit significant gravitational stratification and lower densities in loop apex regions then lead to longer coronal radiative cooling times and consequently smaller $\delta$. The peak emission measure arises at the peak density, which occurs approximately at the transition from conductive to radiative cooling after the ablative upflows have ceased transporting material into the corona. The emission measure slope coolward of the peak is then dominated by strands that are cooling by radiation and enthalpy-driven losses. For shorter loops $\delta \approx 2$ and the loop cools significantly more quickly than it drains. When the strand temperature reaches 1~MK the density hasn't fallen too far from its peak. Conversely, for longer loops $\delta \approx 1$ and the loop cools about as quickly as it drains. When the strand temperature reaches 1~MK a greater fraction of the material will have been lost through draining than in the short loop case, with a commensurately greater change in density. Since $EM(T) \propto n^2$ we can see that a greater density change in long loops during the radiative and enthalpy-driven cooling phase will lead to a correspondingly steeper emission measure than in short loops in the range 1~MK to $T_{\mbox{peak}}$.

Runs~$15-39$ were designed to show the effect of varying the peak heating rate on the emission measure slope and for this group we have chosen a range of volumetric heating rates such that $0.03 \le E_{H0} \le 0.5$~erg~cm$^{-3}$~s$^{-1}$. The weak trend that emerges is for somewhat steeper slopes with increased heating, though it is by no means universal in either the model or forward modeled sets of slopes. It is clear from our results that merely changing the heating rate is not sufficient to obtain emission measure slopes that are in better agreement with the observed slopes, and beyond some upper limit the predicted peak emission measure temperature will exceed the temperature of peak emission observed from active region cores.

The heating timescales that we have chosen for our study lie in the range $10 \le \tau_H \le 2000$~s. $\tau_H$ is the key parameter because we are aiming to use the emission measure slope as a diagnostic of the heating timescale. Runs~$3-5$, $9-12$, $23-30$, and $40-45$, show the results of varying this parameter. Runs~$3-5$, $9-12$, and $40-45$, show $\alpha_{\mbox{model}}$ generally increasing with $\tau_H$. There is no clear pattern for $\alpha_{\mbox{observed}}$, but we note that where particularly steep slopes are found there can be considerable scatter in the Pottasch emission measure values and a correspondingly small $R^2$ obtained from the linear fit ($R^2$  is the square of the correlation coefficient obtained from standard linear regression). The scatter indicates that the density dependence of the contribution function might be an important factor for these emission measures. Runs~$23-30$ (e.g. pairs [23,27], [24,28], etc.), for which group the loop length is significantly longer, show an overall decrease in $\alpha_{\mbox{model}}$ with increasing $\tau_H$. We see a tendency for $\alpha_{\mbox{observed}}$ to increase with $\tau_H$ (except for the weakly heated pair [23,27] indicating a density-related issue) in Runs~$23-30$.

We can combine the parameters $E_{H0}$ and $\tau_H$ into a single parameter: the total volumetric energy input~/~heating rate $E_H$~erg~cm$^{-3}$. There is a clear relationship between $2L$, $E_H$ and $\log_{10}~T_{\mbox{peak}}$ in Table~\ref{table1}. Longer loops require a weaker total volumetric energy input for consistency with the observed range of $\log_{10}~T_{\mbox{peak}}$, as may be inferred from the general coronal loop scaling laws \citep{Rosner_1978}. There is some evidence in the case of shorter loops to suggest that, for Runs with the same (or similar) $E_H$, stronger heating on a shorter timescale yields a steeper slope than weaker heating on a longer timescale. For example, consider the pairs [4,6], [5,7], and the single Run~8 which has very strong heating on a very short timescale but yields a slope at the lower limit of the observed range. The shallow slopes for loops of length 40~Mm (Runs~$15-18$ and $31-33$) which are heated on the relatively long timescale of 500~s are also indicative. There is no evidence of such a relationship for longer loops. Hence, for short loops our experiments hint that strong heating may be required for consistency with the observed range of $\alpha$ and $\log_{10}~T_{\mbox{peak}}$, which in turn requires a short heating timescale to prevent $\log_{10}~T_{\mbox{peak}}$ from overshooting this range. Therefore if one is able to constrain $E_H$ and $E_{H0}$, from the loop length and $\log_{10}~T_{\mbox{peak}}$, then it may also be possible to constrain $\tau_H$.

Finally, we can examine the effect of the temporal profile of the energy release on the emission measure slope. We choose two profiles: a square~/~top-hat; and a triangular distribution. To draw out the effects of the shape of the temporal profile we compare pairs of Runs in the groups $15-26$ and $31-39$ with the same (or similar) $E_H$. Considering the pairs [16,31], [17,32], [20,34], [21,35], [24,37], and [25,38], we find reasonablly good agreement between $\alpha_{\mbox{model}}$ and $\alpha_{\mbox{observed}}$ in each comparison and no apparent dependence on the temporal profile. While on the one hand it may be a pity that there are no obvious clues to the temporal profile using this diagnostic technique, since it is also an unknown in our modeling then it is reassuring to note that our predictions of the emission measure slope are not strongly dependent on the form that we choose. For a given $E_H$ then a
square or triangular profile should yield a reasonably consistent slope.

In general, we conclude that any reasonable combination of loop and heating properties can explain only the lower end of the range of observed emission measure slopes if the measurements are error free. However, the sizable uncertainties found by \cite{Guennou_2012} imply that low-frequency nanoflares might be consistent with a majority of active region cores or with none at all.  If the loop length and the temperature of peak emission can be measured, and if the slope uncertainties can be reduced, then it should be possible to constrain the total volumetric energy input $E_H$ on each strand. In this case our numerical results show that the emission measure-slope diagnostic may be sufficiently sensitive to constrain $\tau_H$ for loops of $2L \le 40$~Mm.

\section{Summary and conclusions}
\label{summary}

We have run a large number of numerical experiments in order to explore an extensive parameter space of loop and heating properties so that we may determine what the emission measure-slope diagnostic can tell us about the time-dependence of active region core heating. We have found that low-frequency nanoflares, where individual magnetic strands cool fully before being re-energized, are consistent with only the lower end ($1.70 \le \alpha \le 2.60$) of the range of slopes reported from observations ($1.70 \le \alpha \le 5.17$). These observed slopes have sizable uncertainties, however. $\alpha$ exhibits a fairly strong dependence on the loop length which can be understood in the low-frequency nanoflare case because its value is determined by the strands undergoing radiative and enthalpy-driven cooling, where the relationship between $T$ and $n$ has a strong dependence on length \citep{Bradshaw_2010}. In the range
consistent with low-frequency nanoflares we have found that the emission measure-slope diagnostic may be sensitive enough to yield information about the time-dependence of the heating mechanism in the case of short ($2L \le 40$~Mm) loops.

Considering our results for $\alpha_{\mbox{observed}}$ alone, where we have found $0.81 \le \alpha_{\mbox{observed}} \le 2.56$, then Table~\ref{table3b} indicates that approximately 36\% of observed active region cores are consistent with heating by low-frequency nanoflares, assuming no errors in the slope measurements. The uncertainties determined by \cite{Guennou_2012} suggest that as many as 77\% of cores may be consistent with low-frequency nanoflares or as few as none.

We can place further physical constraints on our results and restrict the parameter space by considering the nature of the heating mechanism. If we assume that the heating arises from the
impulsive release of energy from the magnetic field then $E_H$ must be limited to the amount of free energy available in the field. The free magnetic energy density is given by:

\begin{equation}
E_B = \frac{1}{8\pi} \left( \epsilon B_p \right)^2,
\label{eqn1}
\end{equation}

\noindent where $B_p$ is the potential component of the magnetic field and $\epsilon$ is a parameter that indicates the level of stress such that $B_s = \epsilon B_p$ is the stress component. A typical value is $\epsilon = 0.3$, with an upper limit of 0.5 \citep{Dahlburg_2005}. By setting $E_B = E_H$ we can determine the magnetic field strength $B$ needed to provide the total volumetric heating, where $B^2 = B_p^2 + B_s^2$. We must bear in mind that not all of the free energy may be released from the field at one time, which means that the magnetic field strength estimated from Equation~\ref{eqn1} may be considered a lower limit.

\citet{Mandrini_2000} studied how the average values of $B$ and $B^2$ along a field line depend on the field line length, $2L$. Table~\ref{BvsL} gives the results for several observed active regions based on equation 9 of their paper. The active regions (numbers 1-6 in the paper) cover a range of age and complexity. Only field lines with photospheric field strengths between 100 and 500 G were used to derive the results. Table~\ref{BvsL} lists $<B^2>^{1/2}$ for the four loop lengths used in our nanoflare simulations. Values are given for the weakest and strongest active region and the average of the 6.

\begin{longtable}{c c c c}
\caption{Average magnetic field strength $\left(<\!B^2\!>^{1/2}\right)$ versus loop length.} \\
\hline
$2L$ & $B_{\mbox{min}}$  & $B_{\mbox{max}}$ & $B_{\mbox{avg}}$ \\
 $[$Mm] & [G] & [G] & [G] \\
\hline
\endhead
20 & 127 & 211 & 167 \\
40 & 83  & 189 & 136 \\
80 & 42  & 150 &  94 \\
160 & 18 & 89 &  51 \\
\hline
\label{BvsL}
\end{longtable}

Table~\ref{table4} shows a subset of the experiments from our study that most closely match observations and also satisfy the magnetic energy constraint. The criteria for selection are that $6.45 \le \log_{10} T_{\mbox{peak}} \le 6.75$, $\alpha_{\mbox{observed}} \ge 1.7$ and $B < B_{\mbox{avg}}$, where $B_{\mbox{avg}}$ is the length-dependent value in the last column of Table~\ref{BvsL}. Regular font indicates that the most restrictive energy constraint $\epsilon=0.3$ satisfies these criteria, and italics indicate that only the less restrictive constraint $\epsilon=0.5$ can satisfy these criteria. The Runs are listed in order of increasing $\alpha_{\mbox{observed}}$. Of the original 14 Runs from \cite{Bradshaw_2011}, only [8,11,12,13,14] are eligible for inclusion in Table~\ref{table4}. We can see that the maximum value of $E_H$ must be somewhere in the region of 50~erg~cm$^{-3}$ to avoid unphysical values of $B$, though exceptions are possible (e.g., Run~18), especially in magnetically strong active regions (third column of Table~\ref{BvsL}). In the case of short loops ($\le 20$~Mm) this limit indicates strong volumetric heating (to reach $6.45 \le \log_{10} T_{\mbox{peak}} \le 6.75$) and short timescales (to satisfy $E_H \le 50$~erg~cm$^{-3}$). Run~8 provides a good example. For longer loops weaker volumetric heating over extended timescales, in comparison with shorter loops, yields emission measure peaks, observed slopes and magnetic field strengths in the right range.

\begin{longtable}{c c c c c c c}
\caption{The Subset of Numerical Experiments that Most Closely Match Observations.}\\
\hline
Run~\# & $2L$ & $E_H$ & $\alpha_{\mbox{model}}$ & $\alpha_{\mbox{observed}}$ & $B(\epsilon=0.3)$ & $B(\epsilon=0.5)$ \\
 & [Mm] & [erg~cm$^{-3}$] & & & [G] & [G] \\
\hline
\endhead
8  & 20 & 50 & 1.79 & 1.73 & 123 & 79 \\
{\it37} & {\it160} & {\it15} & {\it1.98} & \it{1.76} & {\it68} & {\it43} \\
{\it24} & {\it160} & {\it15} & {\it2.01} & {\it1.77} & {\it68} & {\it43} \\
{\it14} & {\it80} & {\it30} & {\it2.01} & {\it1.80} & {\it96} & {\it61} \\
{\it36} & {\it80} & {\it50} & {\it1.81} & {\it1.82} & {\it123} & {\it79} \\
{\it12} & {\it80} & {\it30} & {\it1.90} & {\it1.85} & {\it96} & {\it61} \\
19 & 80 & 7.5 & 1.40 & 1.85 & 48 & 31 \\
{\it44} & {\it80} & {\it30} & {\it2.05} & {\it1.90} & {\it96} & {\it61} \\
{\it18} & {\it40} & {\it125} & {\it1.95} & {\it1.93} & {\it195} & {\it125} \\
42 & 80 & 6.0 & 1.91 & 1.93 & 43 & 27 \\
13 & 80 & 10 & 1.90 & 2.12 & 55 & 35 \\
20 & 80 & 15 & 1.53 & 2.16 & 68 & 43 \\
34 & 80 & 15 & 1.81 & 2.16 & 68 & 43 \\
{\it45} & {\it80} & {\it60} & {\it2.04} & {\it2.16} & {\it135} & {\it87} \\
11 & 80 & 10 & 2.25 & 2.22 & 55 & 35 \\
43 & 80 & 12 & 1.95 & 2.28 & 60 & 39 \\
23 & 160 & 7.5 & 1.68 & 2.47 & 48 & 31 \\
41 & 80 & 3.0 & 1.46 & 2.56 & 30 & 19 \\
\hline \label{table4}
\end{longtable}

In summary; we can use the loop length and the magnetic field strength to constrain $E_H$ and, of the components of $E_H$, we can use $T_{\mbox{peak}}$ to constrain $E_{H0}$ and then $E_H / E_{H0}$ yields $\tau_H$. Our studies show that the emission measure slope is essentially independent of the temporal envelope chosen for the heating and so $\tau_H$ is the key parameter. For low-frequency nanoflares we find predicted slopes in the range $\alpha \le 2.6$.

Finally, we can place limits on the slope that can be obtained in the case of heating by low-frequency nanoflares. The emission measure slope coolward of the peak is determined by the strands undergoing radiative cooling and enthalpy-driven draining. The emission measure in a particular temperature range depends upon two factors: (a) the density of the strands as they cool through that range; and (b) the number of strands in that range. A characteristic emission measure can be calculated by weighting the density-dependent emission measure by the number of strands. Since the number of strands in a particular temperature range depends upon how quickly the plasma cools through that range, $\Delta \tau(T)$, we can write

\begin{equation}
\left<EM(T)\right> = \frac{1}{\tau_{\mbox{cool}}} \int EM(T) \Delta \tau(T),
\label{eqn2}
\end{equation}

\noindent where $\tau_{\mbox{cool}}$ is the total cooling time. We know that

\begin{equation}
EM(T) = n^2 D,
\label{eqn3}
\end{equation}

\noindent where $D$ is the line-of-sight depth of the emitting plasma, and

\begin{equation}
\Delta \tau(T) \propto \frac{P}{n^2
\Lambda(T)}~\mbox{where}~\Lambda(T) = \chi T^b. \label{eqn4}
\end{equation}

\noindent To address the first of the above factors, (a), we require an expression for the density in terms of the temperature of the cooling strands. Following \cite{Bradshaw_2010} we can write

\begin{equation}
T \propto n^\delta~\mbox{where}~\delta = (\gamma-1) + \frac{\tau_V}{\tau_{CR}},
\label{eqn5}
\end{equation}

\noindent where $\gamma = 5/3$, $\tau_V$ is the coronal draining timescale and $\tau_{CR}$ is the coronal radiative loss timescale. We then have

\begin{equation}
n \propto T^\frac{1}{\delta},
\label{eqn6}
\end{equation}

\noindent and since $EM(T) \propto n^2$ we can write

\begin{equation}
EM(T) \propto T^\frac{2}{\delta}.
\label{eqn7}
\end{equation}

\noindent To address the second factor, (b), we require an expression for the cooling timescale in terms of the temperature. Using Equation~\ref{eqn4} we can see that

\begin{equation}
\Delta \tau(T) \propto \frac{T^{1-b}}{n}
\label{eqn8}
\end{equation}

\noindent and substituting Equation~\ref{eqn6} gives

\begin{equation}
\Delta \tau(T) \propto T^{1-b-\frac{1}{\delta}}.
\label{eqn9}
\end{equation}

\noindent We can see from Equations~\ref{eqn2}, \ref{eqn7} and \ref{eqn9} that

\[
\left<EM(T)\right> \propto T^{\frac{2}{\delta}} T^{1-b-\frac{1}{\delta}}
\]

\noindent and so

\begin{equation}
\left<EM(T)\right> \propto T^\alpha~\mbox{where}~\alpha=\frac{1}{\delta}+1-b.
\label{eqn10}
\end{equation}

\noindent The slope of the emission measure, $\alpha$, in any temperature range then depends upon the parameter $\delta$ and the slope of the radiative loss function $b$.

We can place an absolute upper limit on $\alpha$ by recalling that in the limit of pure enthalpy-driven cooling ($\tau_V \rightarrow 0$ and negligible corona radiative losses) $\delta = \gamma-1 = 2/3$ and

\begin{equation}
\alpha_{\mbox{max}} = \frac{5}{2} - b.
\label{eqn11}
\end{equation}

\noindent Typical values of $b$ in the temperature range over which emission measure slopes are calculated are given by $b=-1/2$ \citep{Bradshaw_2010} or $b=-3/2$ \citep{Klimchuk_2008}. These yield $\alpha(b=-1/2)_{\mbox{max}} = 3.0$ and $\alpha(b=-3/2)_{\mbox{max}} = 4.0$. \cite{Bradshaw_2010} showed that $\delta$ decreases with increasing loop length and to approach $\delta=2/3$ would require an unfeasibly long loop for our study. We are limited to loops of approximately 160~Mm in total length to be consistent with active region core scales. Table~2 of \cite{Bradshaw_2010} shows that $\delta(L=160~{\mbox{Mm}}) \approx 1.50$, which yields $\alpha(b=-1/2)=2.17$ and $\alpha(b=-3/2)=3.12$.
In the case of shorter loops Table~2 of \cite{Bradshaw_2010} shows that $\delta \approx 2.00$, which yields $\alpha(b=-1/2)=2.00$ and $\alpha(b=-3/2)=3.00$. We note that \cite{Sturrock_1990} performed a related analysis and found that $E\!M(T) \propto T^{1-b}$.

In consequence, we conclude that it is not possible for the low-frequency nanoflare scenario to yield emission measure slopes coolward of the peak that exceed these values of $\alpha$, depending upon the radiative loss function. In the case of $b=-1/2(-3/2)$, approaching $\alpha=3(4)$ would require loops of unphysical length for active region cores. This explains why our experiments are limited to slopes of about 2.6, because the maximum length in our study is 160~Mm ($2.00 \le \alpha \le 3.00$ for the values of $b$ considered). Therefore, we determine that low-frequency nanoflares cannot explain emission measure slopes in active region cores where $\alpha > 3.00$.

A partial solution to the difficulty of obtaining sufficiently steep emission measure slopes with low-frequency nanoflares might be found from the super-position of coronal loops. Consider a scenario in which a number of particularly long loops ($2L > 160$~Mm, for example) cross the active region core along the observer's line-of-sight, but have foot-points located outside the core. The contribution to the emission measure of such loops might serve to raise the limit on $\alpha$ due to their smaller values of $\delta$, though they are still bound by the lower limit $\delta=2/3$ which yields $\alpha_{\mbox{max}}$ less than the steepest slopes found by \cite{Warren_2012} ($\alpha = 4.50$) and \cite{Schmelz_2012} ($\alpha=5.17$). One might argue that the higher densities of short loops (shorter than a gravitational scale height) means they would make the dominant contribution to the line-of-sight emission and hence to $\alpha$, but their radiative lifetimes are also short meaning that fewer strands contribute to the emission in a given temperature range (equation~\ref{eqn2}). Consequently, we might expect long loops and short loops to have similar emission measures. We can resolve the question of which loops make the dominant contribution by noting that density is strongly dependent on length, and length only enters the weighted emission measure via the parameter $\delta$. We have already demonstrated that $\alpha_{\mbox{max}}$ is smaller for shorter loops and restricted to an upper limit, that lies below many observationally measured values, for longer loops.

In practical terms, then, it seems that to find slopes greater than 2.6 for current radiative loss calculations we must appeal to some other heating scenario. Low-frequency nanoflares are one limit, where the time between heating events on a single strand (the re-energization timescale) is longer than the cooling~/~draining timescale. The opposite limit is steady heating, where the strand has no time to cool before the onset of the next heating event. In order to steepen the slope it is necessary to increase the amount of hot material in the region of $T_{\mbox{peak}}$ relative to the amount of material near 1~MK. In the next paper of this series we will investigate repeating nanoflares on a single strand, where the re-energization timescale is less than the cooling~/~draining timescale, in order to determine whether this mechanism can maintain enough hot material, relative to cooler material, to explain the steepest observed slopes (e.g. $\alpha > 3$).

\acknowledgements

SJB and JAK acknowledge support for this work by the NASA SR\&T program. We thank the International Space Science Institute (ISSI) for hosting the International Team led by SJB and Helen Mason, and the team members for the fruitful discussions that took place during the meeting held there in February 2012. We also thank the referee for their comments and suggestions.


\end{document}